\documentclass[reprint, superscriptaddress, amsmath,amssymb, aps, prl]{revtex4-1}

\usepackage[colorlinks=true,linkcolor=blue]{hyperref}
\usepackage{natbib}
\usepackage{graphicx}% Include figure files
\usepackage{dcolumn}% Align table columns on decimal point
\usepackage{bm}% bold math

\begin{document}
%\preprint{APS/123-QED}
\title{Boundary-induced phase in epitaxial iron layers}

% ************ NAMES ************
\author{Anna L. Ravensburg}
\affiliation{Department of Physics and Astronomy, Uppsala University, Box 516, 75120 Uppsala, Sweden}

\author{Miros\l{}aw Werwi\'{n}ski}
\affiliation{Institute of Molecular Physics, Polish Academy of Sciences, ul. M. Smoluchowskiego 17, 60-179 Pozna\'{n}, Poland}

\author{Justyna Rych\l{}y-Gruszecka}
\affiliation{Institute of Molecular Physics, Polish Academy of Sciences, ul. M. Smoluchowskiego 17, 60-179 Pozna\'{n}, Poland}

\author{Justyn Snarski-Adamski}
\affiliation{Institute of Molecular Physics, Polish Academy of Sciences, ul. M. Smoluchowskiego 17, 60-179 Pozna\'{n}, Poland}

\author{Anna Elsukova}
\affiliation{Thin Film Physics Division, Department of Physics, Chemistry and Biology (IFM), Linköping University, 58183 Linköping, Sweden}

\author{Per O. Å. Persson}
\affiliation{Thin Film Physics Division, Department of Physics, Chemistry and Biology (IFM), Linköping University, 58183 Linköping, Sweden}

\author{J\'{a}n Rusz}
\affiliation{Department of Physics and Astronomy, Uppsala University, Box 516, 75120 Uppsala, Sweden}

\author{Rimantas Brucas}
\affiliation{Department of Materials Science and Engineering, Uppsala University, Box 35, 75103 Uppsala, Sweden}

\author{Bj\"orgvin Hj\"ovarsson}
\affiliation{Department of Physics and Astronomy, Uppsala University, Box 516, 75120 Uppsala, Sweden}

\author{Peter Svedlindh}
\affiliation{Department of Materials Science and Engineering, Uppsala University, Box 35, 75103 Uppsala, Sweden}

\author{Gunnar K. P\'{a}lsson}
\affiliation{Department of Physics and Astronomy, Uppsala University, Box 516, 75120 Uppsala, Sweden}

\author{Vassilios Kapaklis}
\affiliation{Department of Physics and Astronomy, Uppsala University, Box 516, 75120 Uppsala, Sweden}

% ************ DATE ************
%\date{\today}

% ************ ABSTRACT ************
\begin{abstract}

We report the discovery of a boundary-induced body-centered tetragonal ({\it bct}) iron phase in thin films deposited on MgAl$_{2}$O$_{4}$~($001$) substrates. We present evidence for this phase using detailed x-ray analysis and ab-initio density functional theory calculations. A lower magnetic moment and a rotation of the easy magnetisation direction are observed, as compared to body-centered cubic ({\it bcc}) iron. Our findings expand the range of known crystal and magnetic phases of iron, providing valuable insights for the development of heterostructure devices using ultra-thin iron layers.

\end{abstract}

\maketitle

% ************ Introduction ************
The magnetic properties of iron are multifaceted.
This is reflected in the results obtained from investigations on the electric control of magnetic domains \cite{Multistate_Fe_Switching}, magnetic anisotropy \cite{Elmers1990, Allenspach1992, Metoki1993}, magnetic damping \cite{Khodadadi2020}, as well as magnetic interface effects \cite{Urano1988, Balogh2013, Ibrahim2022}.
Not only single layers of Fe are of relevance, Fe in multilayers and superlattices such as Fe/Cr \cite{Etienne1988, Parkin1990}, Fe/V \cite{hjorvarsson}, Fe/Au \cite{Fuss1992}, Fe/MgO \cite{Faure-Vincent2002, Moubah2016, Magnus2018}, or Fe/MgAl$_{2}$O$_{4}$ \cite{Sukegawa2010, Miura2012, Belmoubarik2016, Masuda2017, Xiang2018} exhibits non-trivial properties.
The epitaxial matching of the layers is of particular importance \cite{Martin2007, Sander1999}, since strain and crystal structure can have large effects on the magnetic properties \cite{Moruzzi1989, Andrieu1995_PRB, Friak2001, Burkert2004}.
For example, iron can be ferromagnetic, low-spin or high-spin, antiferromagnetic or even non-magnetic \cite{Moruzzi1989, Andrieu1995_PRB, Friak2001}, all depending on its tetragonal distortion ($c/a$) and unit cell volume. 
Therefore, access to unstrained ultra thin Fe layers is of large importance to enable the separation of boundary and strain effects. 

Recently, it was found that Fe~($001$) layers can be epitaxially grown on single crystalline MgAl$_{2}$O$_{4}$~($001$) substrates \cite{Lee:2017kj, Khodadadi2020, Ravensburg2022}.
A 45~degree in-plane rotation of the Fe unit cell, relative to the unit cell of the substrate, provides growth conditions with an epitaxial misfit of only -0.2\% compared to bulk {\it bcc} Fe~($001$) \cite{Hosseini2008, Ganesh2013, Sukegawa2010, Ravensburg2022}.
Thus, the tetragonal distortion is expected to be low, and the Fe film is structurally similar to {\it bcc} Fe.
Consequently, the crystal quality of Fe~($001$) layers can be significantly improved, as compared to Fe layers grown on MgO~($001$) \cite{Meyrheim_Fe_MgO, Khodadadi2020, Ravensburg2022} or Al$_2$O$_3$~($11\bar{2}0$) \cite{Muehge1994}.
This opens up new alternatives to investigate the effects of layer thickness on, e.g., the magnetic properties of Fe, with only minor substrate-induced strain effects, which is explored in this Letter.% for ultra thin Fe(001) layers.

% ************ Results and discussion ************
% ---------------
\begin{figure}[t!] 
\centering
\includegraphics{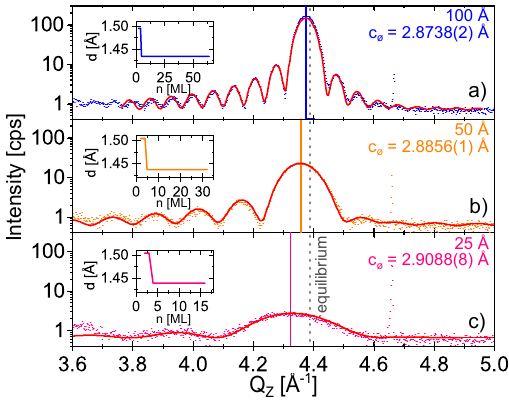}
\caption{X-ray diffraction patterns around the Fe~($002$) Bragg peak of a) 100~{\AA}, b) 50~{\AA}, c) 25~{\AA} thick Fe films grown on MgAl$_{2}$O$_{4}$~($001$). Fits are shown as red lines. The corresponding out-of-plane lattice parameter $c_\text{\o}$ is shown on the right. Insets: Evolution of the out-of-plane atomic distance $d$ as a function of the number of atomic monolayers $n$.} 
\label{fig:XRRD}\hfill
\end{figure}
% ---------------
The Fe films were deposited using direct current magnetron sputtering, with nominal thicknesses $t_\text{Fe}$ in the range of 6--100~{\AA}, at a substrate temperature of 619(2)~K.
All films were capped at ambient temperature with either Pd, Pt, or Al$_{2}$O$_{3}$.
The purpose of the capping is to protect the Fe layer from oxidation.
Measuring identical Fe layers with different capping layers allowed us to explore the effect of the outer boundary on the investigated properties. 
Representative x-ray diffraction patterns using Cu K$_{\alpha 1}$ radiation around the specular Fe~($002$) Bragg peak of three samples with 25, 50, and 100~{\AA} Fe layer thickness are displayed in Fig.~\ref{fig:XRRD}.
As seen in the figure, the peak intensity, position, and shape are different for these samples.
The intensity is found to increase quadratically with increasing thickness, as expected for fully structurally coherent layers \cite{Fewster1996,Birkholz2005}.
Furthermore, the positions of the Fe~($002$) Bragg peak are shifted towards smaller angles with decreasing thickness of the layers.
The out-of-plane atomic distance $d$ in Fe layers of 25, 50, and 100~{\AA} thickness is elongated compared to equilibrium {\it bcc} Fe \cite{Vassent1996, LandoltBornstein1994}, with the average out-of-plane lattice parameters $c_\text{\o}$ along [$001$] being 2.9088(8), 2.8856(1), and 2.8738(2)~{\AA}, respectively.
Hence, the average out-of-plane lattice parameter is consistent with a tetragonal distortion, which appears to increase with decreasing film thickness \cite{Hosseini2008, Ganesh2013, Sukegawa2010, Ravensburg2022}.
The full-width-at-half-maximum (FWHM) of the Fe~($002$) rocking curves are all below 0.04~degrees, independent of layer thickness, consistent with near-perfect single crystalline growth of all these layers.
An atomic registry of the interface between Fe and the substrate was confirmed with atomic-resolved high-angular annular dark field scanning transmission electron microscopy imaging.
Furthermore, no evidence for any oxidation or structural damage was found near the interface (see supplemental material (SM)).

The observation of Laue oscillations around the Fe~($002$) Bragg peak (see Fig.~\ref{fig:XRRD}) for all three samples \cite{Ying2009, Ravensburg2022, Ravensburg2023Fit} provides additional information on the structural coherency.
The asymmetry in the intensity of the Laue oscillations around the ($002$) peak is consistent with the presence of a change in out-of-plane inter-planar atomic spacing in the Fe layers \cite{Vartanyants2000, Robinson2001, Miller2022, Ravensburg2023Fit}.
To obtain information on the shape of the profile of the inter-planar spacing, we performed simulations of the Bragg peak and the Laue oscillations (see SM) using \textsc{GenL} \cite{Ravensburg2023Fit}.
The results of the fitting illustrated in Fig.~\ref{fig:XRRD} are consistent with the presence of two distinct, coherently scattering regions in all samples: 3--4 monolayers closest to the substrate with large tetragonal out-of-plane distortion ($d=1.50$~{\AA}), while the rest of the films has a lattice parameter close to unstrained Fe.

% ---------------
\begin{figure}[t!] 
\centering
\includegraphics{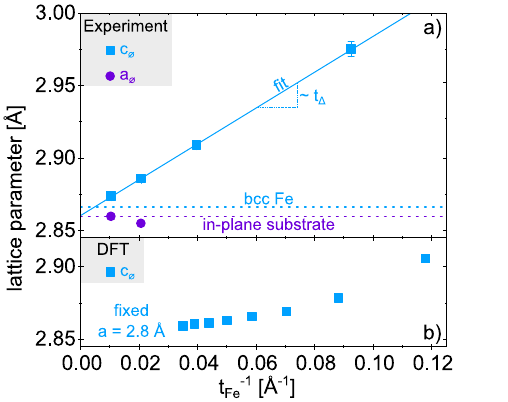}
\caption{a) Experimental in-plane $a_\text{\o}$ and out-of-plane $c_\text{\o}$ lattice parameters for epitaxial Fe films with different layer thicknesses $t_{\mathrm{Fe}}$. The Fe layers are capped with Pd. The choice of capping material did not affect the results, see SM for details. The dashed lines correspond to the equilibrium values of {\it bcc} Fe \cite{Vassent1996, LandoltBornstein1994} and the MgAl$_{2}$O$_{4}$ substrate, while the solid line shows the fit of Eq.~\ref{equ:c_av} to the data. b) DFT calculated average out-of-plane lattice parameter $c_\text{\o}$ for a fixed in-plane lattice parameters $a = 2.8$~{\AA} plotted over inverse Fe layer thickness $t_{\mathrm{Fe}}^{-1}$.} 
\label{fig:experimental}\hfill
\end{figure}
% ---------------

In Fig.~\ref{fig:experimental}a we illustrate a summary of the results from the structural analysis, including asymmetric reflections, namely the Fe~($002$) and ($112$) Bragg peaks (see SM for details).
As seen in the figure, the average out-of-plane lattice parameter $c_\text{\o}$ scales with the inverse thickness of the layers.
The results were fitted using, 
% ---------------
\begin{equation}
\label{equ:c_av}
c_\text{\o} = c_\text{1} \left(1 - \frac{t_\text{$\Delta$}}{t_\text{Fe}}\right) + c_\text{2}\frac{t_\text{$\Delta$}}{t_\text{Fe}},
\end{equation}
% ---------------
where $c_\text{1}$ and $c_\text{2}$ denote the out-of-plane lattice parameters of the two regions and  $t_\text{$\Delta$}$ denotes the extension of the region closest to the substrate.
The intercept of the y axis corresponds to infinitely thick Fe layers, with negligible contribution from the interface region.
From the fitting we get $c_\text{1}=2.862(5)$~{\AA}, which is within the uncertainty identical to the unstrained lattice parameter of {\it bcc} Fe (2.866~{\AA}) \cite{Vassent1996}.
The slope of the fit in Fig.~\ref{fig:experimental}a is proportional to the thickness of the interface layers $t_\text{$\Delta$}$ and the difference in the lattice parameters of the two regions, $(c_2 - c_1) t_\text{$\Delta$} = 0.58(5)$~{\AA}$^2$.
The corresponding values of $(c_2 - c_1) t_\text{$\Delta$}$ from the fitting of the diffraction data (see Fig.~\ref{fig:XRRD}) are 0.56, 0.79, and 0.56~{\AA}$^{2}$ for the 100, 50, and 25~{\AA} thick layers, respectively.
The extension of the interface layer needs to be smaller than 11~{\AA}, as the data point from that sample is captured by the model in Fig.~\ref{fig:experimental}.
Thus, the analysis of the diffraction data and the modelling of the shift of the (002) peak are consistent.

For the samples with 100 and 50~{\AA} thick Fe layers, the average in-plane lattice parameter $a_\text{\o}$ was determined to be 2.860(2) and 2.856(3)~{\AA}, respectively, which closely matches the substrate (2.859~{\AA}) \cite{LandoltBornstein2004}.
Consequently, if the tetragonal distortion of the interface region originated from an elastic response to the biaxial strain, this would correspond to 0.91 in Poisson's ratio equivalent, i.e., the ratio of transverse to longitudinal extension strain, which is not physical for isotropic materials \cite{Wojciechowski2003, Huang2012}.

We used density functional theory (DFT) calculations to explore the contribution of finite-size effects on the obtained results.
Consequently, first the calculations were performed on free-standing Fe layers.
The total energy for a tetragonally distorted {\it bct} structure was found to be lower than that obtained for {\it bcc} when the thickness was below 9 monolayers (see SM).
For biaxially clamped Fe layers with $c/a > 1$, corresponding to the same strain state as experimentally determined for Fe on MgAl$_{2}$O$_{4}$~($001$), the calculated average out-of-plane lattice parameter $c_\text{\o}$ is plotted as a function of inverse Fe layer thickness in Fig.~\ref{fig:experimental}b.
A profound tetragonal distortion is obtained for layers in the few monolayer limit, while the obtained effect is thickness-dependent.
The effect decreases with increasing film thickness, giving rise to a change in slope, as seen in Fig.~\ref{fig:experimental}b.
Considering the size of the calculated effect and the observed changes with thickness, we conclude that the contribution from finite size to the experimentally found structural distortion is small, as compared to the interface effect described above. 

\begin{figure}[t!] 
\centering
\includegraphics{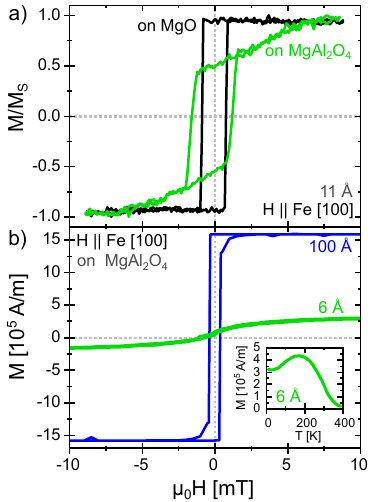}
\caption{Easy axis magnetic hysteresis of a) thin films of 11~\AA~Fe grown on MgAl$_{2}$O$_{4}$~($001$) and MgO~($001$), capped with 50~{\AA} Pd and measured using a longitudinal magneto-optical Kerr effect (L-MOKE) setup and b) 6 and 100~\AA~Fe grown on MgAl$_{2}$O$_{4}$~($001$) capped with 50~{\AA} Al$_2$O$_3$ measured in a superconducting quantum interference device (SQUID). Inset: 
Magnetisation of the 6~{\AA} Fe layer at a constant applied field of 5~mT as a function of temperature.} 
\label{fig:magnetism-main}\hfill
\end{figure}

Having established the presence of an interface layer with a deviating lattice parameter, we now turn our attention to its effects on the magnetic properties.
Hysteresis curves that were measured with an applied field along the bulk Fe in-plane magnetic easy axis direction, i.e., Fe~[$100$] of 11~{\AA} Fe layers deposited simultaneously on MgAl$_{2}$O$_{4}$~($001$) and MgO~($001$), are shown in Fig.~\ref{fig:magnetism-main}a.
A square-shaped hysteresis curve is observed for the Fe layer on MgO~($001$), typical for an easy axis magnetisation loop, in stark contrast to the Fe layer deposited on MgAl$_{2}$O$_{4}$~($001$).
Measurements with applied field along the bulk Fe in-plane magnetic hard axis direction, i.e., Fe~[$110$], confirm a hard axis behaviour for the sample on MgO and an increased remanence for the sample on MgAl$_{2}$O$_{4}$ (see SM for details).
These observations are consistent with  an exchange of easy and hard axes when Fe is grown on MgAl$_{2}$O$_{4}$~($001$) and MgO~($001$) substrates, respectively.
Measurements of the out-of-plane component of the magnetisation excluded any out-of-plane contribution to the magnetisation.

% ---------------
% ---------------

The contribution of the interface region to the overall magnetic properties can also be inferred by comparing samples with vastly different thicknesses.
For instance, Fig.~\ref{fig:magnetism-main}b displays hysteresis curves measured along the Fe~[$100$] direction for a 6 and 100~{\AA} Fe layer deposited on MgAl$_{2}$O$_{4}$~($001$).
We observe a reduction in the saturation magnetisation with decreasing thickness, accompanied also by a distinct change in the shape of the hysteresis curve.
The temperature dependent magnetisation of the 6~{\AA} Fe layer was determined in an applied magnetic field of 5~mT along the Fe~[$100$] direction.
An initial increase with decreasing temperature is observed, followed by a decrease with decreasing temperature below 150~K.
These observations are consistent with an increase in magnetic anisotropy, exceeding the torque provided by the external field at about 150~K, in line with the conclusions above.
The tetragonal distortion at the interface is not the only factor affecting the magnetic properties.
For example, Fe forms Fe-O bonds at the MgAl$_{2}$O$_{4}$~($001$) interface \cite{Meyrheim_Fe_MgO, Sukegawa2010}, which can contribute to lower magnetisation, a change in the magnetic anisotropy, as well as lower Curie temperature.
For the 6~{\AA} Fe layer, the contribution of finite-size effects on the ordering temperature is non-negligible and can be determined by empirical models, as seen in \citet{Zhang:2001db} and \citet{BHPRL}. 

% ************ Conclusions ************
In summary, we have identified a boundary-induced state in Fe at the interface with a MgAl$_{2}$O$_{4}$~($001$) substrate.
The interface state in Fe is argued to result from an electronic proximity effect with the substrate.
The easy axis of the observed {\it bct} phase is rotated by $\pi$/4 as compared to bulk Fe, which gives rise to changes in the obtained anisotropy with thickness.
These findings add to the understanding and optimal design of ML thick Fe layers in heterostructures such as Fe/MgO or Fe/MgAl$_{2}$O$_{4}$, which hold a potential for magnetic tunnel junctions and future three-dimensional memory storage devices.

% ************ Acknowledgements ************
VK and PS acknowledge financial support from the Swedish Research Council (Project No. 2019-03581 and 2021-04658). GKP acknowledges funding from the Swedish Energy Agency (Project No. 2020-005212). MW, JRG, and JSA acknowledge financial support from the National Science Center Poland under decisions DEC-2018/30/E/ST3/00267 (SONATA-BIS 8) and DEC-2019/35/O/ST5/02980 (PRELUDIUM-BIS 1). The authors also acknowledge access to the Swedish National Infrastructure for Advanced Electron Microscopy, ARTEMI, supported by the Swedish Research Council (VR) and The Foundation for Strategic Research (SSF), through grants 2021-00171 and RIF21-0026. JR acknowledges financial support from the Swedish Research Council (2021-03848), Olle Engkvist's foundation (214-0331) and Knut and Alice Wallenberg foundation (2022.0079).

The data that support the findings of this study are available from the authors upon reasonable request.

% ************ BIB ************
%\bibliographystyle{apsrev4-2}
%\bibliography{FeGrowth}
\providecommand{\noopsort}[1]{}\providecommand{\singleletter}[1]{#1}%
%

% ************ SUPPLEMENTAL ************
%%%%%%%%%% Merge with supplemental materials %%%%%%%%%%
\pagebreak
\onecolumngrid
\newpage
\begin{center}
\textbf{\large Supplemental Material: Boundary-induced phase in epitaxial iron layers}
\end{center}
%%%%%%%%%% Merge with supplemental materials %%%%%%%%%%
%%%%%%%%%% Prefix a "S" to all equations, figures, tables and reset the counter %%%%%%%%%%
\setcounter{equation}{0}
\setcounter{figure}{0}
\setcounter{table}{0}
\setcounter{page}{1}
\makeatletter
\renewcommand{\theequation}{S\arabic{equation}}
\renewcommand{\figurename}{Supplementary FIG.}
\renewcommand{\thefigure}{{\bf \arabic{figure}}}
\renewcommand{\bibnumfmt}[1]{[S#1]}
\renewcommand{\citenumfont}[1]{S#1}
\renewcommand{\thepage}{S-\arabic{page}}
%%%%%%%%%% Prefix a "S" to all equations, figures, tables and reset the counter %%%%%%%%%%

% ************ METHODS ************
\section{Methods}

% ************ Growth ************
\subsection{Growth details}

Fe thin films were deposited using direct current (dc) magnetron sputtering on single crystalline MgAl$_{2}$O$_{4}$~($001$) substrates of size 10$\times$10~mm$^2$ at floating potential.
The substrates were annealed at 1273(2)~K in vacuum for 600~s prior to growth.
The base pressure of the growth chamber was below 5$\times$10$^{-7}$~Pa.
In order to prevent surface oxidation of the films due chemical affinity of Fe for oxygen \cite{Campbell1997}, the samples were capped at ambient temperature ($<$~313(2)~K) with Pd and selected samples with Pt or Al$_{2}$O$_{3}$ instead.
The capping layer thickness was varied between 10~{\AA} and 50~{\AA}.
The depositions were carried out at a power of 50~W in an Ar atmosphere (gas purity $\geq$~99.999~\%, and a secondary getter based purification) from elemental Fe (0.67~Pa Ar, dc), Pd (1.07~Pa Ar, dc), and Pt (1.07~Pa Ar, dc) targets and an Al$_{2}$O$_{3}$ (0.67~Pa Ar, rf) compound target.
The targets were sputter cleaned against closed shutters for 60~s prior to each deposition.
The target-to-substrate distance in the deposition chamber was around 0.2~m.
The deposition rates (Fe: 0.1~{\AA}/s, Pd: 0.6~{\AA}/s, Pt: 0.8~{\AA}/s, Al$_{2}$O$_{3}$: 0.01~{\AA}/s) were calibrated prior to the growth.
Details on the optimization of the deposition temperature can be found elsewhere \cite{Ravensburg2022-2}.
In order to ensure thickness uniformity, the substrate holder was rotated at 30~rpm during deposition.

% ************ XRR & XRD ************
\subsection{X-ray scattering details}

X-ray reflectometry (XRR) and diffraction (XRD) were carried out in a Bede D1 diffractometer equipped with a Cu x-ray source.
The setup includes a G\"obel mirror and a 2-bounce-crystal on the incidence side.
Furthermore, a circular mask with a diameter of 5~mm, an incidence and a detector slit, 0.5~mm each, were used.
The scattered x-rays were detected with a Bede EDRc x-ray detector.
The instrument angles were aligned to the sample surface for XRR and to the Fe crystal planes for XRD measurements. 
The measured XRR data was fitted using \textsc{GenX} \cite{Bjorck2007, Glavic_Bjorck_2022} enabling the determination of layer thickness and roughness and of the scattering length density (SLD) profile.
However, twinning in the substrates, and therefore also in the epitaxially growing Fe layers, may lead to an overestimation of the layer roughnesses.
In diffraction, the samples were measured with a combination of coupled $2\theta$-$\theta$ and rocking curve scans.
Incident and reflected angles were converted to momentum transfer $Q$.
Peak positions in $Q_\text{z}$ (out-of-plane component) and the width of rocking curves were determined by fitting with a Gaussian and a Lorentzian profile, respectively.
All error bars for fits of scattering data are statistical and do not include systematic errors arising from alignment or absorption.
X-ray diffraction patterns including Laue oscillations were fitted with \textsc{GenL} \cite{Ravensburg2023Fit-2}.
The fitted parameters relating to the Fe layer were: the average number of coherently scattering planes contributing to the Laue oscillations $N_\text{L}$ and the average out-of-plane atomic distance $d_\text{002}$.
Moreover, a strain profile was fitted within the model to account for strain relaxation \cite{Ravensburg2023Fit-2}.
A contribution of other factors to the asymmetry, e.g., the presence of finite atomic terraces or crystal grains \cite{Miller2022-2} was omitted.
Hence, as the employed model takes into account only strain as a source for the asymmetry in the Laue oscillations, the change in out-of-plane atomic distance over layer thickness might be overestimated.

Reciprocal space mapping (RSM) around the ($002$) and ($112$)~Fe Bragg reflections was carried out in a Bruker D8 Discover equipped with a Cu x-ray source and a 1D Bruker LynxEye detector.
The setup included a G\"obel mirror, a 0.2~mm slit, and a Soller slit on the incidence side as well as a 4.6~mm slit, a beam collimator, and a Soller slit on the detector side.
The maps were fitted with Gaussian and Lorentzian intensity profiles in $Q_\text{z}$ (out-of-plane component) and $Q_\text{x}$ (in-plane component) directions, respectively.
Out-of-plane and in-plane lattice parameters were extracted according to \cite{Birch1995}.
In addition, the accuracy of the determined parameters was confirmed by repeating the measurements with a sample of 100~{\AA} Fe layer thickness on the above mentioned  K$_{\alpha1}$ monochromatic Bede D1 x-ray machine.

The Poisson's ratio equivalent $\nu$ was calculated based on the relation between in-plane and out-of-plane distortion of a unit cell due to uniaxial stress \cite{Birkholz2005-2}:
% ---------------
\begin{equation}
\label{equ:Poisson_lat}
\frac{2\nu}{1+\nu} = \frac{1-\frac{a_\text{eq}}{c}}{\frac{c-a}{a}},
\end{equation}
% ---------------
where $a_\text{eq}$, $a$, and $c$ denote the equilibrium, the strained in-plane, and the strained out-of-plane lattice parameters, respectively.

% ************ Magnetization ************
\subsection{Details on measurements of magnetization}
Magnetization measurements were performed at ambient temperature using a longitudinal (L-MOKE) and polar (P-MOKE) magneto-optical Kerr effect setup with $s$-polarized light.
The magnetic response was measured parallel and perpendicular to an in-plane applied magnetic field for L- and P-MOKE, respectively.
The data was averaged over 10 full loop recordings.
For L-MOKE, the samples were probed along the Fe~[$100$] and [$110$] directions, hence, along the expected magnetic easy and hard axis, respectively.
Selected samples were additionally probed at azimuthal angles between 350 and 190~degrees for the determination of their normalized remanent magnetization.
Displayed error bars correspond to statistical errors only.

The magnetic moment of selected samples was probed in a superconducting quantum interference device (SQUID) at temperatures between 10 and 400~K using a Quantum Design Magnetic Property Measurement System.
The magnetic field in these measurements was applied in-plane along the Fe [$100$] direction.

% ************ TEM ************
\subsection{Details on transmission electron microscopy}

Lattice- and atomic-resolved high-angular annular dark field scanning transmission electron microscopy (HAADF-STEM) imaging were carried out at ambient temperature with double-Cs corrected FEI Titan3 60–300, operated at 300~kV.
The samples for TEM analysis were prepared by ion milling in a Gatan precision ion miller after fine mechanical polishing.

% ************ DFT ************
\subsection{Details on computations}

Density functional theory (DFT) calculations were performed using the full-potential local-orbital scheme (FPLO18.00-52)~\cite{koepernik_full-potential_1999,eschrig_chapter_2004}, employing the generalized gradient approximation (GGA) in the Perdew-Burke-Ernzerhof parametrization~\cite{perdew_generalized_1996}.
The calculations were performed on {\it bcc} Fe model systems with a ($001$) surface and thicknesses ranging from one to twenty monolayers. 
Preserving the geometry of the square lattice basis, the lattice parameter $a$ of the basis was optimized (free-standing Fe) or fixed (clamped Fe) while the distances in the $z$ direction between atomic monolayers were optimized using interatomic forces.

% ************ RESULTS ************
\section{Results}

% ************ XRR ************
\subsection{Layering}

X-ray reflectivity curves for Fe layers grown on MgAl$_{2}$O$_{4}$~($001$) with a nominal thickness of 25, 50, and 100~{\AA} are displayed in Supplementary Fig.~\ref{fig:xrr}.
The scattering length density profiles corresponding to the displayed reflectivity fits are shown as insets in the respective color.
The profiles agree well with the intended layering.
The Fe layer thickness of the three thin films displayed in Supplementary Fig.~\ref{fig:xrr} are determined to be 25(1), 48(1), and 96(1)~{\AA}.
As they lie within 4\% of the intended layer thickness, we refer to the nominal layer thickness throughout the text.
The Fe layer roughness of the 25, 50, and 100~{\AA} thick Fe layers are 1(1), 1(1), and 3(1)~{\AA}, respectively.
% ---------------
\begin{figure}[h!] 
\centering
\includegraphics{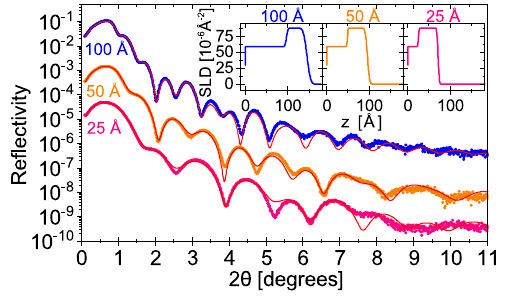}
\caption{X-ray reflectivity of Fe films of varying thicknesses. Fits are shown as red lines. The curves have been vertically shifted for clarity. The respective scattering length density profiles are shown as insets.}
\label{fig:xrr}\hfill
\end{figure}
% ---------------

% ************ Laue oscillations ************
\subsection{Simulations and fittings of Laue oscillations}

XRD patterns of Fe~($001$) with different out-of-plane atomic spacing profiles including Laue oscillations have been simulated using \textsc{GenL} \cite{Ravensburg2023Fit-2}.
The results for two 100~{\AA} Fe films are displayed in Supplementary Fig.~\ref{fig:sim}, the first consisting of 11~{\AA} {\it bct} Fe with out-of-plane $d = 1.5$~{\AA} and 89~{\AA} equilibrium {\it bcc} Fe $d = 1.433$~{\AA} and the second film consisting of 100~{\AA} equilibrium {\it bcc} Fe.
From the simulations it is evident that the periodicity of the Laue oscillations is affected by these structural differences, because the spacing between the oscillations around the Fe~($002$) Bragg peak corresponds to the thickness of the {\it bcc} Fe layer.
Hence, the spacing between two maxima is larger corresponding to smaller real space distances for the {\it bct}/{\it bcc} Fe sample.
Furthermore, the oscillations become asymmetric around the Bragg peak, decaying faster on the high angle side, what has also been observed in experimentally measured x-ray diffraction patterns for Fe thin films deposited on MgAl$_{2}$O$_{4}$~($001$), as displayed for a 100~{\AA} Fe layer in Supplementary Fig.~\ref{fig:sim_fit}.
% ---------------
\begin{figure*}[h!] 
\centering
\includegraphics[width=\textwidth]{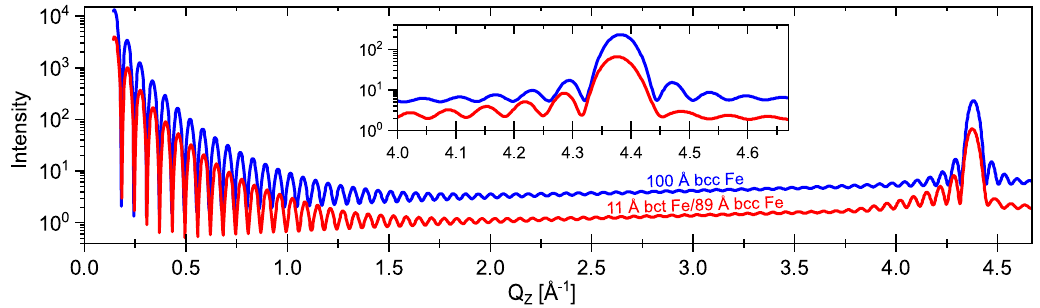}
\caption{X-ray diffraction pattern simulations using \textsc{GenL} \cite{Ravensburg2023Fit-2} for 100~{\AA} Fe layers consisting of, first, 11~{\AA} {\it bct} Fe with $d = 1.5$~{\AA} and 89~{\AA} equilibrium {\it bcc} Fe (red) and second, 100~{\AA} equilibrium {\it bcc} Fe (blue). A zoom onto the region around the Fe~($002$) Bragg peak is displayed in the inset.}
\label{fig:sim}\hfill
\end{figure*}
% ---------------

The diffraction data in Supplementary Fig.~\ref{fig:sim_fit} was fitted using \textsc{GenL} \cite{Ravensburg2023Fit-2} and employing an exponentially decaying strain profile as well as the profile described above, having a {\it bct} Fe/{\it bcc} Fe bilayer out-of-plane interplanar spacing.
It is evident that the {\it bct} Fe/{\it bcc} Fe bilayer profile captures the faster decay of the oscillations on the high angle side in more detail.
As this is valid for all samples within this study, the fit employing the bilayer profile was used for the analysis.
Based on the {\it bct} Fe/{\it bcc} Fe bilayer fits (displayed in the main text), the thickness of the {\it bct} Fe layer is about 3, 4, and 3~ML for the samples with 100, 50, and 25~{\AA} Fe layer thickness, having an out-of-plane interplanar spacing of 1.495, 1.504, 1.504~{\AA}, respectively.
94, 93, and 89~\% of the 100, 50, and 25~{\AA} samples contribute to coherent scattering, respectively.
% ---------------
\begin{figure*}[h!] 
\centering
\includegraphics{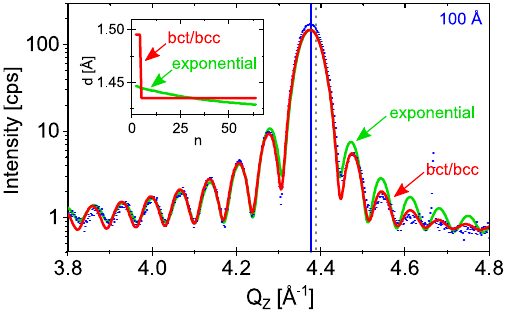}
\caption{X-ray diffraction pattern of a 100~{\AA} Fe layer on MgAl$_{2}$O$_{4}$~($001$). Fits using \textsc{GenL} \cite{Ravensburg2023Fit-2} are displayed as green and red lines for an exponentially decaying strain profile and a {\it bct}/{\it bcc} grown Fe film, respectively. The respective strain profiles are shown in the inset. The vertical line marks the Bragg peak position. The dotted line indicates the peak position expected for equilibrium {\it bcc} Fe.}
\label{fig:sim_fit}\hfill
\end{figure*}
% ---------------

% ************ RSMs ************
% ---------------
\begin{figure*}[t!] 
\centering
\includegraphics[width= \textwidth]{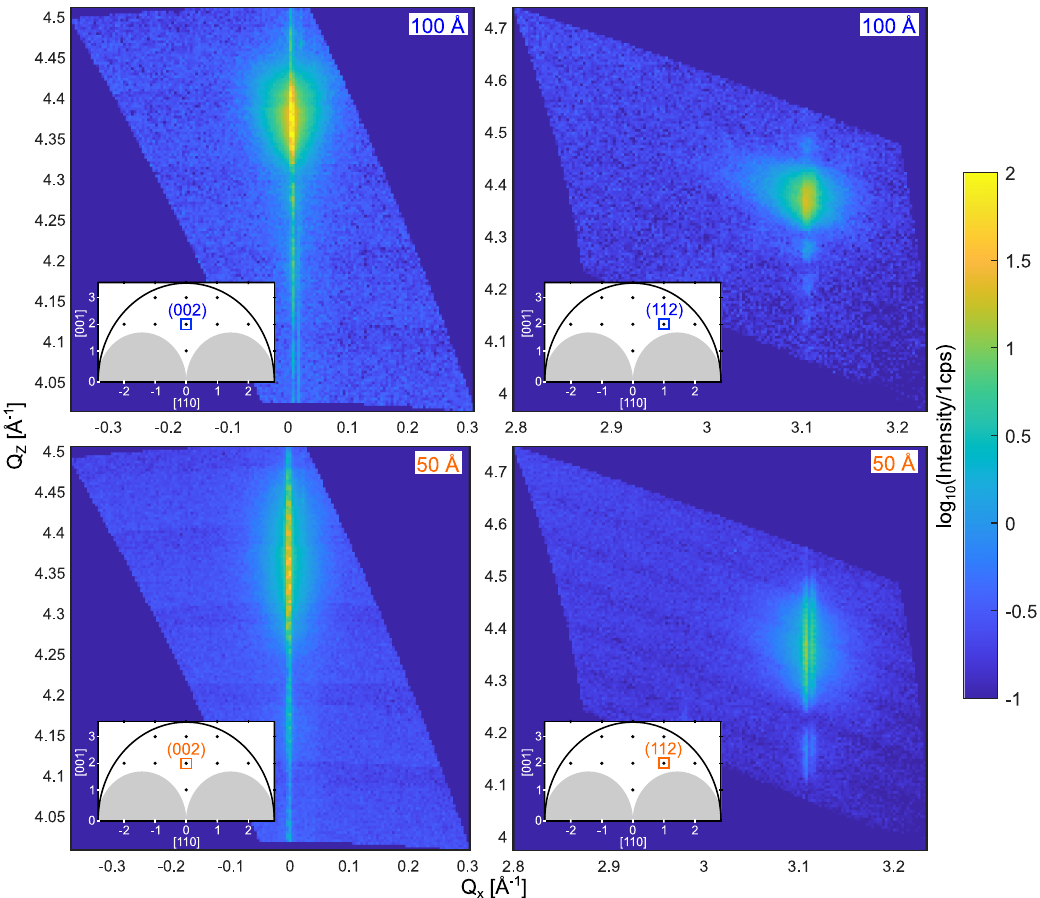}
\caption{Reciprocal space maps measured around the Fe~($002$) and ($112$) Bragg reflections on samples with 100~{\AA} (top) and 50~{\AA} (bottom) Fe layer thickness. Schematic illustrations of the position of the respective peak in reciprocal space are shown as insets.} 
\label{fig:rsm}\hfill
\end{figure*}
% ---------------
\subsection{Reciprocal space maps}

The reciprocal space maps (RSMs) for samples with 100 and 50~{\AA} Fe layer thickness grown on MgAl$_{2}$O$_{4}$~($001$) are displayed in the top and bottom panel of Supplementary Fig.~\ref{fig:rsm}, respectively.
RSMs of samples with thicknesses below 50~{\AA} Fe layer thickness could not be measured as the Bragg peak intensity was determined to lie below the detector background level of the used diffractometer.
Sharp Bragg peaks are observable.
The average out-of-plane lattice parameters $c_\text{\o}$ based on the RSMs deviate by less than 0.11\% from the previously determined values based on specular scans.
The corresponding in-plane lattice parameter $a_\text{\o}$ were determined to be 2.860(2) and 2.856(3)~{\AA} for the samples with 100 and 50~{\AA} thick Fe layers, respectively.
The size of the in-plane lattice parameter for the 100~{\AA} thick Fe layer based on fitting of the RSM was confirmed with TEM measurements in diffraction mode.
Laue oscillations are observable for both samples for measurements along $Q_\text{z}$.
The spacing of the oscillations is larger for thinner Fe layers.
While the peak shape is rather broad along $Q_\text{z}$, it is observed to be significantly more narrow along $Q_\text{x}$, in agreement with the observed small FWHM of the rocking curves for both samples.
However, along $Q_\text{z}$ two parallel narrow lines are observable corresponding to different crystal twins.
Twinning in epitaxial thin films often results from substrate twinning, which is common for this type of substrates \cite{Schroeder2015}.

% ************ TEM study ************
\subsection{Real space imaging}
Lattice- and atomic-resolved high-angular annular dark field scanning transmission electron microscopy (HAADF-STEM) results from the 100~{\AA} Fe sample is displayed in Supplementary Fig.~\ref{fig:TEM}.
Epitaxial registry of Fe~($001$) and MgAl$_{2}$O$_{4}$~($001$) is clearly observed.
The atoms at the interfaces grow in registry with low defect density in the distinct layers, whereby the substrate/Fe interface is sharper with improved layering compared to the Fe/Pd interface.
Hence, x-ray diffraction and STEM measurements confirm almost defect-free growth of Fe~($001$) on MgAl$_{2}$O$_{4}$~($001$), in line with the reported critical thickness of 785~{\AA} of Fe~($001$) on MgAl$_{2}$O$_{4}$~($001$) \cite{Ravensburg2022-2}.
% ---------------
\begin{figure}[t!] 
\centering
\includegraphics{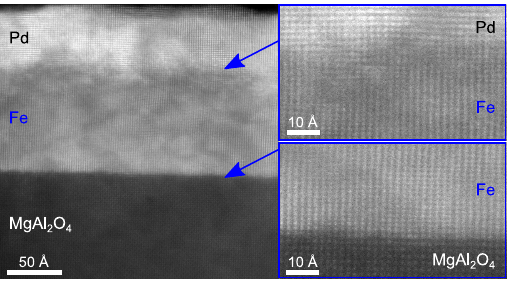}
\caption{Left: Cross-section HAADF-STEM image of a MgAl$_{2}$O$_{4}$/100~{\AA} Fe/50~{\AA} Pd film. Right: Magnifications on the Fe/Pd interface (top) and the substrate/Fe interface (bottom).} 
\label{fig:TEM}\hfill
\end{figure}
% ---------------

% ************ Capping study ************
\subsection{Study of the influence of the capping layer}

A possible physical origin of the observed tetragonal distortion of Fe on MgAl$_{2}$O$_{4}$~($001$) might be a change of spin-orbit coupling in Fe layers at the thin film limit, due to the strong correlation of the structural and magnetic properties of Fe \cite{Moruzzi1989-2, Andrieu1995_PRB-2, Friak2001-2}.
One possible cause for a change in spin-orbit coupling for thin Fe layers might be the large susceptibility of the 50~{\AA} Pd capping layer \cite{Hase2014}.
Hence, the influence of the capping layer on the Fe average out-of-plane lattice parameter $c_\text{\o}$ was investigated and the results are displayed in Supplementary Fig.~\ref{fig:cap}.
First, 25~{\AA} Fe thin films were grown on MgAl$_{2}$O$_{4}$~($001$) with a variable Pd capping layer thickness of 10, 20, and 50~{\AA}.
Second, 100~{\AA} Fe thin films were grown on MgAl$_{2}$O$_{4}$~($001$) with a Pt capping layer thickness of 50~{\AA}.
Finally, 25, 50, and 100~{\AA} thick Fe layers were grown on MgAl$_{2}$O$_{4}$~($001$) with a 25~{\AA} thick Al$_2$O$_3$ capping layer.
The Fe out-of-plane lattice parameter $c_\text{\o}$ of all samples shows the same linear increase with inverse Fe layer thickness independent of the choice of capping layer material or thickness.
Therefore, an influence of the capping layer on the spin-orbit coupling of Fe layers at the thin film limit is excluded.
% ---------------
\begin{figure}[h!] 
\centering
\includegraphics{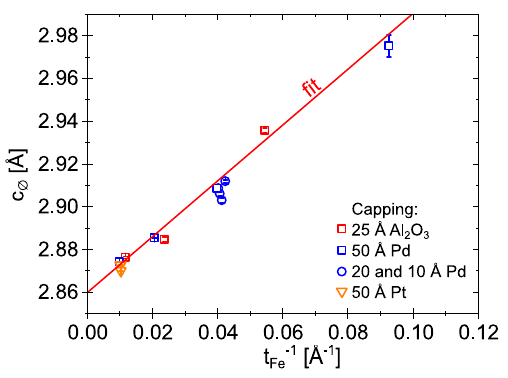}
\caption{Average out-of-plane lattice parameter $c_\text{\o}$ of Fe plotted over inverse Fe layer thickness for samples with different capping layers. A fit is shown in red.} 
\label{fig:cap}\hfill
\end{figure}
% ---------------

% ************ Unit cells ************

\subsection{Results of DFT calculations and schematic illustration of Fe unit cells}
% ---------------
\begin{figure}[h!] 
\centering
\includegraphics{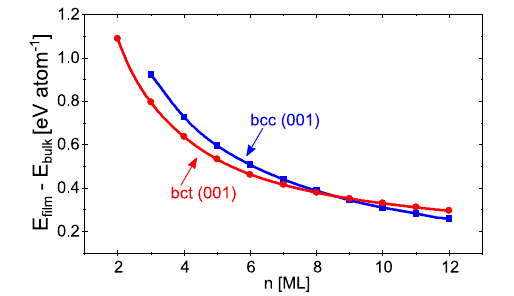}
\caption{Total energies ($E_\text{film}$) of free-standing {\it bcc} and {\it bct} Fe~(001) thin films as a function of number of Fe atomic ML $n$. The results converge to the total energy of the bulk Fe ($E_\text{bulk}$). The curves are guides to the eye.} 
\label{fig:Ecurve}\hfill
\end{figure}
% ---------------
Density functional theory (DFT) calculations at 0~K have indicated that below an Fe layer thickness of 9 ML a {\it bct} crystal structure with an elongated out-of-plane interplanar spacing is energetically more favorable compared to the equilibrium Fe {\it bcc} crystal structure, as can be seen in Supplementary Fig.~\ref{fig:Ecurve}.
The calculations were performed for a [$001$] {\it bct} growth direction of free-standing layers, i.e., in vacuum.
If the Fe lattice parameters $a$ and $c$ are not both relaxed but $a$ is fixed, the calculated total energy is higher.
Schematic illustrations of the Fe {\it bcc} and {\it bct} cells are displayed in Supplementary Fig.~\ref{fig:structures}a and b, respectively.
The observed tetragonally distorted Fe unit cell in thin Fe films grown on MgAl$_{2}$O$_{4}$~($001$) is schematically displayed in Supplementary Fig.~\ref{fig:structures}c for the 11~Å thick Fe thin film.
The dimensions are based on the determined average out-of-plane lattice parameter $c_\text{\o}$.
% ---------------
\begin{figure}[h!] 
\centering
\includegraphics{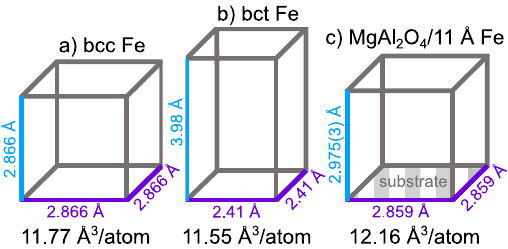}
\caption{Schematic display of Fe unit cells with different crystal structures: a) {\it bcc} Fe equilibrium structure with a cubic lattice parameter of 2.866~{\AA} \cite{Vassent1996-2, LandoltBornstein1994-2}, b) theoretically calculated Fe {\it bct} crystal structure, and c) experimentally determined crystal structure in 11~{\AA} Fe thin films grown on MgAl$_{2}$O$_{4}$~($001$) assuming fully strained epitaxial growth with clamping at the substrate. Unit cell volumes per atom are displayed below each structure.} 
\label{fig:structures}\hfill
\end{figure}
% ---------------
Based on the magnetic phase diagram calculated by \citet{Moruzzi1989-2}, ferromagnetism is expected for the {\it bct} Fe structures observed experimentally and in DFT.
The calculated Fe {\it bct} structure, and the experimentally determined average Fe structure for 11~{\AA} layer thickness have unit cell volumes of 11.55, and 12.16~{\AA}$^3$/atom and $c/a$ ratios of 1.65, and 1.04, respectively and are, hence, well within the ferromagnetic regime \cite{Moruzzi1989-2, Martin2007-2}.

If the Fe is not calculated free-standing but clamped, i.e., having a fixed in-plane lattice parameter $a$, the calculated average out-of-plane lattice parameter $c_\text{\o}$ is shown in Supplementary Fig.~\ref{fig:cav} for $a=$ 2.859, 2.8, and 2.7~{\AA}.
The in-plane lattice parameter $a=$ 2.859~{\AA} corresponds to the in-plane atomic distance in the MgAl$_{2}$O$_{4}$~($001$) substrate at ambient temperature.
Due to thermal expansion, at 0~K this parameter relates however to $c/a<1$.
Nevertheless, for all selected $a$ a profound increase in average out-of-plane lattice parameter $c_\text{\o}$ is observed for thinner films.
% ---------------
\begin{figure}[h!] 
\centering
\includegraphics{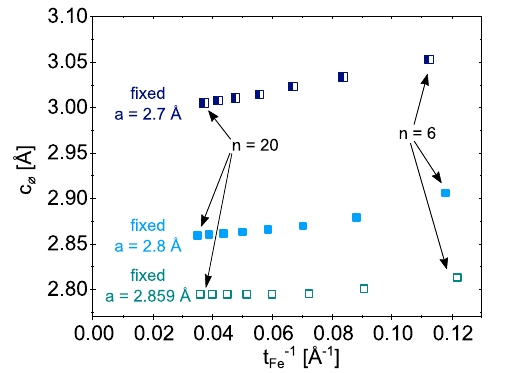}
\caption{DFT calculated average out-of-plane lattice parameter $c_\text{\o}$ for fixed in-plane lattice parameters $a$ plotted over inverse Fe layer thickness.} 
\label{fig:cav}\hfill
\end{figure}
% ---------------

% ************ Magnetic properties ************
\subsection{Measurements of the magnetic properties at ambient temperature}

The magnetic properties of Fe thin films with variable thickness grown on MgAl$_{2}$O$_{4}$~($001$) substrates were investigated in a MOKE and a SQUID setup.
L-MOKE hysteresis loops of samples with 100, 25, and 11~{\AA} Fe layer thickness for applied fields along Fe~[$100$] are displayed in Supplementary Fig.~\ref{fig:magnetism}a, b, and c, respectively.
While the loops for 100 and 25~{\AA} Fe layer thickness are square-shaped, the loop for the sample with 11~{\AA} Fe layer thickness exhibits a rhombic shape with significantly larger coercivity of around 1.6~mT and higher saturation field hinting towards a change in magnetocrystalline anisotropy, in line with results reported in literature for structures with from cubic distorted Fe unit cells.
For example, tetragonally distorted Fe in MgO/Fe/MgO heterostructures exhibits a growth-induced uniaxial in-plane magnetic anisotropy \cite{Park1995}, in contrast to the fourfold in-plane magnetic anisotropy in {\it bcc} Fe originating from spin-orbit coupling in combination with the crystalline symmetry.
L-MOKE measurements of the remanent to saturation magnetization ratio $M_R/M_S$ for samples with 100, 25, and 11~{\AA} Fe layer thickness are shown in Supplementary Fig.~\ref{fig:magnetism}d.
The samples with 25 and 100~{\AA} Fe thickness show both a bulk-like fourfold magnetocrystalline anisotropy.
However, the azimuthal angle-dependent remanent state of the sample with 11~{\AA} Fe layer thickness does not show the same behavior, in particular for applied magnetic fields along the expected Fe easy axes, i.e., Fe~[$100$].

% ---------------
\begin{figure*}[h!] 
\centering
\includegraphics[width= \textwidth]{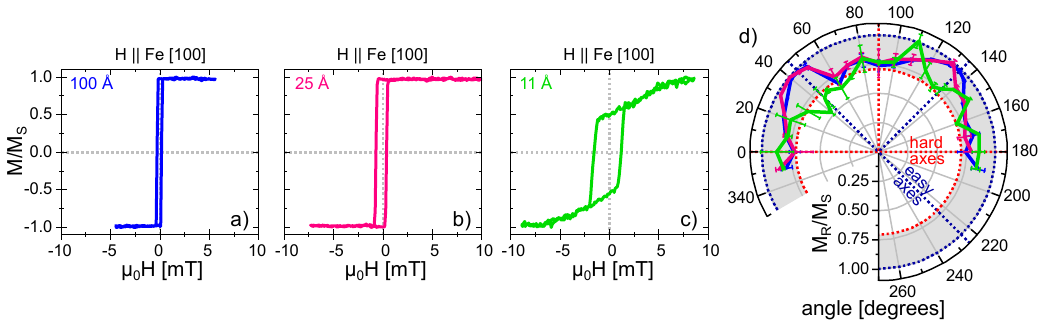}
\caption{a-c) Magnetic hysteresis loops measured with an applied field along [$100$] direction (one of the magnetic easy axis of bulk Fe) on Fe thin films of different layer thickness grown on MgAl$_{2}$O$_{4}$~($001$) substrates. All samples were capped with 50~{\AA} Pd. d) Azimuthal dependence of the remanent magnetization for the same samples with 100, 25, and 11~Å Fe layer thickness. The grey shaded area marks the ratios of $M_R/M_S$~=~$\sqrt{2}/2$ for a field along a hard axis and $M_R/M_S$~=~1 for an easy axis.} 
\label{fig:magnetism}\hfill
\end{figure*}
% ---------------

At remanence, the magnetization is not entirely pointing along the expected Fe magnetic easy axis direction for Fe layers of this thickness.
To exclude that the magnetization starts pointing out-of-plane, p-MOKE and out-of-plane SQUID measurements were performed.
No sign of an out-of-plane magnetization was found in either of the measurements.

The change in magnetocrystalline anisotropy for thin Fe layers grown on MgAl$_{2}$O$_{4}$~($001$) can be observed in comparison to Fe layers grown on MgO~($001$) instead (see Supplementary Fig.~\ref{fig:SQUIDmagnetism}a).
For applied fields along Fe~[$110$] hysteresis loops of 11~{\AA} on both substrates exhibit a similar shape, but the sample on MgAl$_{2}$O$_{4}$~($001$) has a higher relative remanence magnetization $M_R/M_S$, indicating that the magnetic hard axis in this sample corresponds to a different direction.

% ---------------
\begin{figure*}[h!] 
\centering
\includegraphics{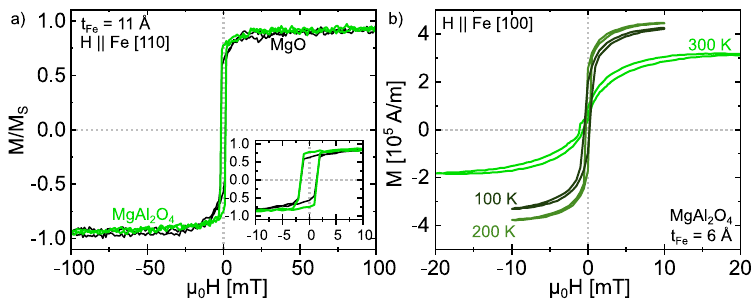}
\caption{Magnetic hysteresis loops a) of 11~{\AA} Fe on MgAl$_{2}$O$_{4}$~($001$) and MgO~($001$) with an applied field along Fe~[$100$].  Both measurements were performed at ambient temperature. b) Temperature dependent magnetic hysteresis loops measured with an applied field along Fe~[$100$] using SQUID on a 6~{\AA} thick Fe thin film grown on MgAl$_{2}$O$_{4}$~($001$). The samples were capped with a) 50~{\AA} Pd and b) 50~{\AA} Al$_2$O$_3$.} 
\label{fig:SQUIDmagnetism}\hfill
\end{figure*}
% ---------------

\subsection{Temperature dependent measurements of the magnetic properties}

The temperature dependence of the measured magnetization of a sample with 6~{\AA} Fe layer thickness is displayed in the main text and revealed a lowered Curie temperature compared to equilibrium {\it bcc} Fe.
The hysteresis loops measured at 100, 200, and 300~K and displayed in Supplementary Fig.~\ref{fig:SQUIDmagnetism}b confirm the observed change in magnetization for fields applied along Fe~[$100$], which at an external field of 5~mT is first increasing with decreasing temperature but showing a maximum at around 150~K.

% ************ BIB ************
%\bibliographystyle{apsrev4-2}
%\bibliography{FeGrowth}

\providecommand{\noopsort}[1]{}\providecommand{\singleletter}[1]{#1}%
%

% ************ END ************
\end{document}